\documentclass[a4paper]{jpconf}
\usepackage{graphicx}
\usepackage{psfrag}
\usepackage[dvips]{epsfig}
\usepackage{dsfont}
\usepackage{mathrsfs}

\DeclareFontFamily{OT1}{rsfs}{}
\DeclareFontShape{OT1}{rsfs}{m}{n}{ <-7> rsfs5 <7-10> rsfs7 <10->
rsfs10}{} \DeclareMathAlphabet{\mycal}{OT1}{rsfs}{m}{n}
\newcommand{\scri}{\mycal{I}}
\newcommand{\bas}{\begin{eqnarray*}}
\newcommand{\eas}{\end{eqnarray*}}
\newcommand{\be}{\begin{equation}}
\newcommand{\ee}{\end{equation}}

\begin{document}
\title{Numerical calculations near spatial infinity}
\author {An\i l Zengino\u{g}lu}
\address{Max-Planck Institut f\"ur Gravitationsphysik (AEI), Am M\"uhlenberg 1, 14476 Potsdam Germany}
\ead{anil@aei.mpg.de}

\begin{abstract}
After describing in short some problems and methods regarding the smoothness of null infinity for isolated systems, I present numerical calculations in which both spatial and null infinity can be studied. The reduced conformal field equations based on the conformal Gauss gauge allow us in spherical symmetry to calculate numerically the entire Schwarzschild-Kruskal spacetime in a smooth way including spacelike, null and timelike infinity and the domain close to the singularity.
\end{abstract}

\section{Motivation}
The main purpose of this article is to study the feasibility of the conformal Gauss gauge in numerical applications. The importance of this gauge lies in its usefulness in the analysis of the gravitational field in a neighbourhood of spatial infinity, as we discuss in the following. For review articles see \cite{Friedrich04,Frauendiener04}.
\subsection{Isolated Systems}
The motivation to study the regularity of null infinity for asymptotically flat spacetimes arises from interest in the following question: \textit{How can we describe the gravitational field of isolated systems in a rigorous and efficient way?}

Isolated systems in general relativity are self-gravitating models of astrophysical sources. Depending on the context, these can be planets, stars, black holes and even galaxy clusters. It is not the source that makes a system isolated, but the asymptotic region. The gravitational field of isolated systems becomes weak as we move away from the source. As nothing is completely isolated from the rest of the universe and as we can not yet test our assumptions by experiment, details of this idealization need to be decided upon in accordance with our physical intuition, expectation and most importantly, in accordance with the equations describing the model. Isolated systems are modeled in general relativity by asymptotically flat spacetimes in which the metric on the complement of a spatially compact set approaches near infinity the flat metric in a specified way. 

Many physical concepts like mass, linear and angular momentum are defined in the asymptotic region with respect to infinity. The use of coordinate dependent limits to infinity for calculating these quantities can deliver misleading results if one is not careful enough. A geometric, coordinate independent way to deal with isolated systems has been proposed by Penrose \cite{Penrose63} by use of conformally rescaled spacetimes with a smooth conformal boundary. The main idea is to adjoin infinity to the physical spacetime by a conformal rescaling of the metric, as is done in complex analysis, where the Riemann sphere is constructed by adjoining infinity to the complex plane \cite{Penrose67}. One studies the equivalence class of the physical metric $\tilde{g}$ given by the conformal rescaling $g=\theta^2 \tilde{g}$ with a function $\theta>0$. As the physical distances go to infinity the conformal factor approaches zero at a specific rate so that the rescaled metric $g$ is regular at the set $\{\theta=0\}$ of the conformal extension if appropriate fall-off conditions are satisfied.  This geometric construction replaces coordinate dependent limits to infinity by local differential geometry at $\{\theta=0\}$ and allows an elegant description of the asymptotic structure of isolated, self-gravitating systems which has proven very fruitful in mathematical analysis.

\begin{figure}[]
  \centering
  \psfrag{i+}{$i^+$}
  \psfrag{Scri}{$\scri^+$}
  \psfrag{i0}{$i^0$}
  \includegraphics[width=0.4\textwidth]{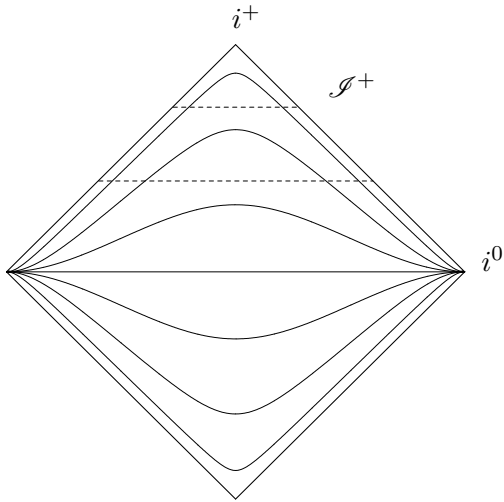}
  \caption{Numerically generated Penrose diagram of the Minkowski spacetime with Cauchy and hyperboloidal surfaces \label{fig:minkowski}}
\end{figure}

The Penrose diagram in Fig.~\ref{fig:minkowski} depicts the causal structure of the Minkowski spacetime illustrating the different asymptotic regions. One can go to infinity along timelike, null or spacelike directions approaching timelike $i^{\pm}$, null $\scri$ or spatial infinity $i^0$. In such a Penrose diagram, light rays are straight lines with 45 degrees to the horizontal. Plotted are also Cauchy surfaces that approach spatial infinity and hyperboloidal surfaces that cut null infinity. The conformal compactification of the Minkowski spacetime, as found for example in \cite{Frauendiener04}, is the simplest application of the idea of conformal compactification.
\subsection{Spatial infinity}
The details of the conformal compactification technique developed by Penrose have been motivated rather by explicit examples or well studied assumptions about the fall-off behaviour of certain fields than a detailed study of the full non-linear Einstein's field equations, which poses the following question \cite{Penrose82}: \textit{How general is the description proposed by Penrose?} 

We would like to know by some general argument whether we have a large class of non-trivial, asymptotically flat, radiative spacetimes that admit a smooth conformal boundary such that we can apply the conformal compactification technique. To answer this question, the solution space to the Einstein equations needs to be studied with an emphasis on the asymptotic structure of the gravitational fields.

The available explicit solutions are not general enough to study the solution space by direct means, therefore we are led to abstract analysis. By sufficient knowledge of the properties of the spacetime, we can get some general results. For example we know that asymptotically flat, vacuum, stationary spacetimes admit an analytic compactification at null infinity \cite{Dain01b}. For more general results that include radiative spacetimes, the initial value problem needs to be studied.

Unfortunately, the analysis of the initial value problem for the Einstein equations in the compactified picture is not straightforward, as the Einstein's equations are not conformally invariant and compactification leads to formally singular equations. However, the equations are conformally regular as Friedrich showed by constructing a system which is equivalent to the Einstein's equations for $\theta>0$ and is regular for all values of the conformal factor so that the equations can be analysed on the conformal extension of the spacetime \cite{Friedrich81}. With this system Friedrich showed that if conformal data on a spacelike hypersurface that cuts null infinity is given such that it is smooth up to null infinity, then the evolution of this data has a smooth conformal boundary  \cite{Friedrich86}. This analysis showed that the decision on the smoothness of null infinity is made at spatial infinity.

By the gluing techniques developed by Corvino and Schoen \cite{Corvino-Schoen} one knows that there exists a large class of non-trivial initial data which is Schwarzschild or Kerr in a neighbourhood of spatial infinity. Combined with Friedrich's earlier results this leads to the existence of a large class of non-trivial radiative vacuum spacetimes with smooth null infinity \cite{Chrusciel-Delay}. 

However, these spacetimes have a special asymptotic structure. The question still remains whether more general spacetimes exist. One would also like to construct such spacetimes numerically. These tasks require a more sophisticated method to study spatial infinity. 
\subsection{Reduced Conformal Field Equations}
A detailed study of the solutions in a neighbourhood of spatial infinity is complicated by the fact that in general point compactification at spatial infinity leads to singular conformal data. However, Friedrich was able to pose a regular finite initial value problem near spatial infinity by using the reduced conformal field equations that he introduced in \cite{Friedrich95}. These equations are based on the conformal Gauss gauge in which spatial infinity in an asymptotically flat spacetime can be represented as a cylinder \cite{Friedrich98}. Subsequent analysis showed that in general, logarithmic singularities arise in a small neighbourhood of spatial infinity. Friedrich obtained necessary regularity conditions on the Bach tensor. Later on, Valiente Kroon showed that these conditions are not sufficient and obtained further obstructions to smoothness of null infinity \cite{ValienteKroon04}. 

While the question about the necessary and sufficient conditions for smoothness of null infinity still remains, we can ask whether one can deal with some mild singular behaviour numerically by using the reduced conformal field equations, which turned out to be the proper tool to study the asymptotic behaviour of solutions in analytic work. A strong interaction between mathematical and numerical studies in this question might give new impulses in both directions. The desire to have numerical simulations which allow one to understand and control the asymptotic structure of spacetimes is the main motivation behind my work.

In the following, we will summarize ingredients of the underlying conformal Gauss gauge for the reduced conformal field equations.
\section{Conformal Gauss gauge}
The main reference for the properties of the conformal Gauss gauge that we discuss below is \cite{Friedrich03}. This gauge is based on conformal geodesics, which are autoparallel curves with respect to a Weyl connection. 
\subsection{The Weyl connection}
The main step from Euclidean geometry to Riemannian geometry is the removal of the assumption of integrability of vectors by parallel transport. A vector parallel transported along a closed curve changes in general its direction due to curvature, while its norm stays the same. Weyl, arguing that the choice of the unit of measurement, i.e. the gauge, should also be subject to local variations, demanded ``the non-integrability of the transference of distances'' \cite{Weyl23}. This simple requirement leads to a rich geometry, which Weyl used in his attempt to unify the gravitational and electromagnetic fields. While his attempt was not successful, the main idea has been the basis during the construction of modern gauge theories \cite{Oraifeartaigh}.

Driven by such conceptual and also other philosophical considerations, Weyl introduced a torsion free connection which is not necessarily the Levi-Civita connection of a metric but preserves the conformal structure and thus satisfies
\be \label{weyl-conn} \hat{\nabla}_{\alpha} g_{\mu\nu} = - 2 f_{\alpha}  g_{\mu\nu}. \ee
If the 1-form $f$ is exact, it can be written as $f = \Omega^{-1} d\Omega$ and the Weyl connection $\hat{\nabla}$ will be the Levi-Civita connection of a metric in the conformal class $\hat{g}_{\mu\nu} = \Omega^{2} g_{\mu\nu}$.  The Weyl connection $\hat{\nabla}$ is invariant under a conformal rescaling $\bar{g}=\vartheta^2 g$ with a function $\vartheta>0$, so the 1-form $f$ transforms according to $\bar{f}=f-\vartheta^{-1}\,d\vartheta$.

The relation of the Weyl connection $\hat{\nabla}$ with a metric connection $\nabla$ of a metric $g$ in the conformal class is given by 
\[ \hat{\nabla} - \nabla = S(f), \qquad {\rm where} \qquad S(f)^{\ \rho}_{\mu \ \nu} \equiv \delta^{\rho}_{\ \mu}f_{\nu} + \delta^{\rho}_{\ \nu}f_{\mu} - g_{\mu\nu} g^{\rho\lambda} f_{\lambda}. \]
We see that the Weyl connections are characterized by 1-forms $f$. Contrary to Weyl who used the additional freedom introduced by the 1-form $f$ to construct a unified field theory, Friedrich exploited it as a gauge freedom of the conformal structure in his construction of the reduced conformal field equations, which are equivalent to the Einstein's field equations. 

A torsionfree connection implies the notion of parallel transport and allows the construction of geodesics, which we will study next. 
\subsection{Conformal geodesics}
Null geodesic congruences, when they are smooth, provide a valuable tool to study the asymptotic and causal structure of spacetimes. As they are invariants of the conformal structure, one might assume that time- or spacelike curves that are conformal invariants might also be useful in such studies, which is indeed the case. Conformal geodesics are conformally invariant in the sense that they are independent of the metric chosen in the conformal class. 

A solution to the conformal geodesic equations does not only provide a spacetime curve, but along the curve also a Weyl connection, a conformal factor and a frame which is orthonormal for a metric in the conformal class. While the equations are independent of coordinates, we will have to write them in some coordinate system, as we will study them in numerical applications. 

Given a metric $g$, the equations that define the conformal geodesic $x^\mu(\tau)$ are written for its tangent vector $\dot{x}^{\mu}(\tau)$ and a covector $f_{\mu}(\tau)$ that defines a Weyl connection
\begin{eqnarray*} 
(\nabla_{\dot{x}}\dot{x})^{\mu}+S(f)^{\ \mu}_{\lambda \ \rho}\dot{x}^{\lambda}\dot{x}^{\rho} &=& 0, \nonumber \\
(\nabla_{\dot{x}}f)_{\mu}- \frac{1}{2}f_{\lambda} S(f)^{\ \lambda}_{\rho \ \mu}\dot{x}^{\rho} &=& L_{\lambda\mu}\dot{x}^{\lambda},
\end{eqnarray*}
where $L_{\mu\nu}=\frac{1}{2}\,R_{ij} - \frac{1}{12}\,R\, g_{ij}$ is the Schouten tensor. Written in terms of the Weyl connection that is defined by the 1-form, $\hat{\nabla} = \nabla + S(f)$, the first equation reads $\hat{\nabla}_{\dot{x}}\dot{x}=0$. The equation for the spacelike frame fields $e_a^{\ \mu}(\tau)$ along the curve is $\hat{\nabla}_{\dot{x}}e_a=0$.

From (\ref{weyl-conn}) we see that on a given curve $x^\mu(\tau)$, the metric $\theta^2 g_{\mu\nu}$ with a conformal factor $\theta>0$ is parallel transported with respect to $\hat{\nabla}$ if $\theta$ satisfies $\hat{\nabla}_{\dot{x}} \theta = \theta \, f_\mu\dot{x}^\mu$ along the curve. A remarkable property of the conformal geodesics is that in a vacuum spacetime the equation for the conformal factor  $\theta(\tau)$ can be solved explicitly, such that $\theta(\tau)$ is known a priori in terms of initial data along the conformal geodesics. For certain initial data, the conformal factor can be written as
\be \label{eq:conf_fac} \theta(\tau) = \frac{\Omega}{\kappa}\left(1-\tau^2\frac{\kappa^2}{\omega^2}\right) \qquad \mathrm{with} \qquad \omega = \frac{2 \ \Omega}{\sqrt{\, h^{ab}\, D_{a}\Omega \, D_b \Omega}}. \ee
$\Omega$ is a conformal factor on the initial hypersurface, $h_{ij}$ is the with $\Omega$ rescaled spatial metric in compactifying coordinates, $\kappa$ is a free function determining the time coordinate at which the conformal geodesic cuts $\scri$. One can also show that
\be \label{eq:dk} d_k := \theta \, f_\mu e^{\ \mu}_{k} = \left(-2 \tau \frac{\kappa \Omega}{\omega^2},\frac{1}{\kappa}e_a(\Omega) \right). \ee  
For a derivation of these results, see \cite{Friedrich95}.

Some of the conformal geodesics leave the spacetime through null infinity. This might seem to be a counter-intuitive behaviour as the curves are timelike everywhere. The explanation is that they experience a non-vanishing acceleration by the 1-form $f$. A similar behaviour show hyperboloidal slices which are spacelike everywhere but also reach null infinity. One should remember that the causal nature of a curve or a surface does not completely determine its asymptotic behaviour. 

To construct a conformal Gauss gauge one uses conformal geodesics in a similar way as one uses metric geodesics to construct the Gauss gauge. One specifies a congruence of timelike vectors, a conformal factor and a 1-form on an initial hypersurface.  The timelike geodesics starting from this surface provide the conformal Gauss coordinates. Spatial coordinates are dragged along. In \cite{Friedrich03} Friedrich constructs conformal Gauss coordinates in the Schwarzschild-Kruskal spacetime and shows that they do not only cover the global solution, but also the conformal extension in a smooth way. We will construct this gauge numerically. 
\subsection{Numerical experiments}
First we describe the general route to construct the gauge on a given background and then we mention the numerical results.

Assume a solution to the Einstein field equations $(\widetilde{\mathcal{M}},\tilde{g})$ has been given. A procedure to construct a conformal Gauss gauge with timelike conformal geodesics and an orthonormal frame along them $( \, x^\mu(\tau), f_\mu(\tau), e_a^{\,\mu}(\tau) \, )$ for a given spacetime solution to the vacuum Einstein equations can be given as follows: 
\begin{enumerate}
\item Find a conformal compactification of the global solution $g=\vartheta^2 \tilde{g}$. In general, you will need different coordinates in different asymptotic regions.
\item On a spatial slice with a spatial metric $\tilde{h}$, introduce compactifying coordinates and rescale $\tilde{h}$ with an appropriate function $\Omega$, such that in these coordinates you have $h=\Omega^2 \tilde{h}$. The metric $h$ is needed for the calculation of the conformal factor (\ref{eq:conf_fac}).
\item Set initial data $(x^\mu(0), \dot{x}^\mu(0), f_\mu(0), e_a^{\,\mu}(0))$ on the initial hypersurface $\tau = 0$ according to 
  \[ f_\mu(0) = \Omega^{-1} \partial_\mu \Omega, \qquad f_\mu(0) \, \dot{x}^\mu(0) = 0, \qquad \theta^2 \tilde{g}(\dot{x},\dot{x})|_{\tau=0} = \frac{\Omega^2}{\kappa^2}\tilde{g}(\dot{x},\dot{x})|_{\tau=0} =  - 1, \]
  \[ \theta^2\, \tilde{g}(e_a, \dot{x})|_{\tau=0} = 0, \qquad \theta^2\,\tilde{g}(e_a, e_a)|_{\tau=0} = 1, \qquad \mathrm{with} \qquad a=1,2,3. \] 
The timelike frame vector is given by $\dot{x}$ itself. The frame is not unique, the freedom corresponds to the freedom of Lorentz rotations. One also needs to choose the free function $\kappa$ that determines the form of $\scri$ in the conformal Gauss gauge.
\item Solve the following system of ordinary differential equations with the above initial data
\begin{eqnarray}\label{congeo-component}
(\partial_\tau x)^\mu &=& \dot{x}^\mu, \nonumber \\
(\partial_\tau\dot{x})^{\mu} &=& - \Gamma_{\lambda\ \rho}^{\ \mu} \dot{x}^\lambda\dot{x}^\rho -2(f_{\nu}\dot{x}^{\nu})\,\dot{x}^{\mu} + (g_{\lambda\rho}\dot{x}^{\lambda}\dot{x}^{\rho})\,g^{\mu\nu}f_{\nu}, \nonumber \\
(\partial_\tau f)_{\mu} &=& \Gamma_{\mu\ \lambda}^{\ \rho} \dot{x}^\lambda f_\rho + (f_{\nu}\dot{x}^{\nu})\, f_{\mu} - \frac{1}{2}(g^{\lambda\rho}f_{\lambda}f_{\rho})\, \dot{x}^{\mu} + L_{\mu\nu}\dot{x}^{\nu}, \nonumber \\
(\partial_\tau e_a)^{\mu} &=& - \Gamma_{\lambda\ \rho}^{\ \mu} \dot{x}^\lambda e_a^{\,\rho} - (f_{\nu} e_a^{\,\nu})\,\dot{x}^\mu - (f_{\nu}\dot{x}^{\nu})e_{a}^{\,\mu} + (g_{\lambda\rho}e_a^{\,\lambda}\dot{x}^{\rho})\, g^{\mu\nu}f_{\nu}.
\end{eqnarray}
\item Check the quality of the solution using (\ref{eq:conf_fac}) and (\ref{eq:dk}). 
\end{enumerate}
The right hand sides of (\ref{congeo-component}) are calculated using the computer algebra package MathTensor. For the integration of the ODE system, I used a 3th order Runge-Kutta integration algorithm. 
\subsubsection{Schwarzschild-Kruskal}
\begin{figure}[htbp]
\centering
\psfrag{T}{\small$\tau$}
\psfrag{i+}{\small$i^+$}
\psfrag{J+}{\small$\scri^+$}
\psfrag{I}{\small$\mathcal{I}$}
\psfrag{d/M}{\small $d/M$}
\psfrag{2}{\tiny 2}
\psfrag{10}{\tiny 10}
\psfrag{300}{\tiny 300}
\psfrag{ingoing}{\small ingoing}
\psfrag{outgoing}{\small outgoing}
\psfrag{r=const}{\small r = const}
\psfrag{singularity}{\small singularity}
\psfrag{horizon}{\small horizon}
\includegraphics[width=0.5\textwidth,height=4cm]{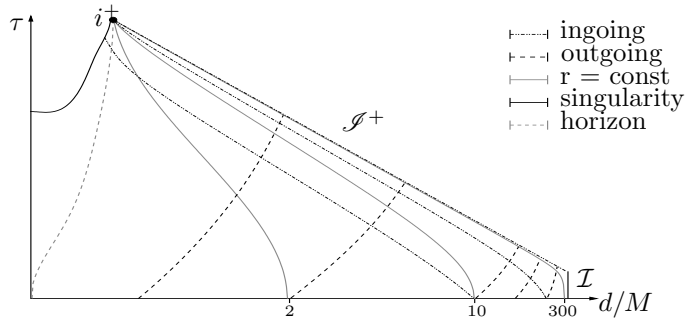}
\caption{Schwarzschild-Kruskal spacetime in a conformal Gauss gauge.\label{fig:conf_geodesics}}
\end{figure}

We calculate the conformal Gauss initial data for the Schwarzschild-Kruskal spacetime. The resulting gauge is illustrated in a numerically generated conformal diagram Fig. \ref{fig:conf_geodesics}. The physical Schwarzschild metric for $r>2m$ reads
\[ \tilde{g}=-\left(1-\frac{2 m}{r}\right) dt^{2} + \left(1-\frac{2 m}{r}\right)^{-1}dr^{2} + r^{2}\,d\sigma^{2}. \]
We transform the metric using the retarded and advanced null coordinates 
\be \label{wv} w = t - (\, r + 2 m \, \ln \,(r-2m\,)), \quad v = t + (r+2m \ \ln \,(r-2m)). \ee
After the inversion $z = \frac{1}{r}$ and the rescaling with $\vartheta=z$ following $g = \vartheta^{2} \tilde{g}$ we get 
\[ g = - z^{2} ( 1-2mz)\,dw^{2} + 2\, dw\, dz + d\sigma^{2}, \quad g = - z^{2} ( 1-2mz)\,dv^{2} - 2\, dv \, dz + d\sigma^{2} \]
The coordinates $w,v$ extend analytically into regions where $r \leq 2 m$. For a simulation through $\scri^+$ we use the retarded null coordinate $w$, for a simulation through the future horizon to the future singularity we use the advanced null coordinate $v$. \\
We give initial data on a $t =$const.~slice. From the transformation (\ref{wv}) we have 
\[ dw = \frac{1}{z^2(1-2mz)}dz, \qquad dv = - \frac{1}{z^2(1-2mz)}dz, \]
which leads to the induced spatial metric
\[ \tilde{h} = \frac{1}{z^4(1-2mz)}\,dz^2 + \frac{1}{z^2}\,d\sigma^2. \]
From now on, we choose $m=2$ and present the calculation in the outgoing case to simplify the formulae. We compactify $\tilde{h}$ using the conformal factor $\Omega=\frac{2z^2}{1-2z}$, so that $h=\Omega^2 \tilde{h}$ is diffeomorph to the standard metric on the three sphere  $h=d\chi^2 + \sin^2\chi\, d\sigma^2$, as can be shown by the transformation $z(\chi)=\frac{\sin\chi}{2(1+\sin\chi)}$.\\
The initial data for the 1-form $f$ can be given as $f(0) = \frac{2(1-z)}{z(1-2z)}dz$. The orhogonality condition $f_\mu \dot{x}^\mu=0$ and the normalisation requirement delivers $\dot{x}(0)= \frac{1-2z}{2 z^2\sqrt{1-4 z}}\kappa\,\partial_w$. The timelike frame vector is given by $e_0 = \dot{x}$. Initial data for the spatial frame vector reads
\[ e_1(0) = \kappa\, \frac{1-2 z}{2 z^2\sqrt{1 - 4 z}}\partial_w + \kappa \, \frac{(1-2z)\sqrt{1- 4 z}}{2}\partial_z. \]

We can also take these steps using other compactifications of Schwarzschild, for example the isotropic compactification. The use of different coordinate systems or different compactifications for calculating the right hand side of (\ref{congeo-component}) does not effect the quality of the calculation. The resulting conformal diagram depends on the choice of the spatial coordinate and the free function $\kappa$. Fig. \ref{fig:conf_geodesics} is taken from \cite{Husa05} and is the result of a numerical calculation where $\kappa$ has been chosen such that $\scri^+$ is a straight line in the corresponding conformal Gauss gauge. Illustrated is the ``upper right part'' of the Penrose diagram for the Schwarzschild-Kruskal spacetime. The lower horizontal line corresponds to the hypersurface $\{t=0\}$ in the standard Schwarzschild coordinates, where also $\{\tau=0\}$. We see that the conformal geodesics cover in a smooth way spacelike, null and timelike infinity and the domain close to the singularity.

A typical question regarding calculations going up to infinity is whether it is not ``too far away''. After all, the Earth where we are doing our measurements is not infinitely far away from the sources and we do not move along null geodesics as an observer along null infinity does. In the Schwarzschild case, we have a length scale $M$ at our disposal, so we can compare the astrophysical situation with the numerical calculations. In the conformal diagram, three curves with constant distances to the event horizon  $d=2 M, 10 M, 300 M$ have been plotted. Numerical codes today have outer boundaries at about $300M$ to $1000M$. The distance $1000M$ corresponds to about 3000km for a two solar mass black hole. Even if we consider supermassive black holes, the corresponding distance is incomparably small with respect to the thousands or millions of light years that separates us from the astrophysical sources. As one sees in the Fig. \ref{fig:conf_geodesics}, already the curve at $300M$ is barely distinguishable from $\scri^+$. In this representation, the numerical effort for simulating the region from $300 M$ to $\scri^+$ is very small whereas in the standard approach putting the outer boundary from $300 M$ to $1000 M$ costs considerable effort in terms of numerical techniques and computational sources. Therefore, one can conclude that the conformal compactification technique is more apt to deal with the asymptotic region.
\subsubsection{Kerr}
Going beyond spherical symmetry, we can solve the conformal geodesic equations in Kerr spacetime. We take the Kerr metric in Boyer-Lindquist coordinates, do the inversion and make an Eddington-Finkelstein-like transformation using ingoing and outgoing light rays as coordinate lines as before. For the outgoing case, the metric becomes
\begin{eqnarray*}
\tilde{g} &=& -\left(1-\frac{2m}{z\Sigma}\right) dw^2 - \frac{2}{z^2} dw\,dz - \frac{4 a m}{z\Sigma}\sin^2\!\theta \,dw\,d\phi \\
 & + & \frac{2a}{z^2}\sin^2\!\theta\,dz\,d\phi+\Sigma\,d\theta^2+\frac{1}{\Sigma}\left(\left(\frac{1}{z^2}+a^2\right)^2- \triangle a^2 \sin^2\!\theta\right) \sin^2\!\theta\, d\phi^2,
\end{eqnarray*}
where
\[ \Sigma = \frac{1}{z^2} + a^2 \cos^2\theta, \qquad \ \triangle = \frac{1}{z^2} - \frac{2m}{z} + a^2. \]
Consider the coordinate transformation $z(\chi)=\frac{\sin \chi}{\delta+m\,\sin\chi}$ with $\delta = \sqrt{m^2-a^2}$. Following \cite{Dain01a}, a $t=$const.~hypersurface in the original Boyer-Lindquist coordinates can be compactified with the conformal factor 
\[ \Omega = \frac{\sin\chi}{\sqrt{\Sigma}}=\frac{\delta z^2}{(1-mz)\sqrt{1+a^2z^2cos^2\theta}}. \]
The compactified spatial metric reads
\[ h = d\chi^2 + \sin^2\chi d\theta^2 + \sin^2\chi \sin^2\theta \left( 1+a^2 \frac{1+2m/(\Sigma\,z(\chi))}{\Sigma \sin^2\chi} \sin^2\chi \sin^2\theta \right) d\phi^2 \]
We get for the initial data from $f=\Omega^{-1}\, d\Omega$
\[ f(0)=\frac{2-mz+a^2z^2\cos^2\!\theta}{z(1-mz)(1+a^2z^2\cos^2\!\theta)}\,dz + \frac{a^2z^2\cos\theta sin\theta}{1+a^2z^2\cos^2\!\theta}\,d\theta, \]
and for $\dot{x}$
\[ \dot{x}(0) = \frac{(1-mz)(1+a^2z^2\cos^2\!\theta)}{\delta z^2 \sqrt{1+a^2z^2\cos^2\!\theta-2mz}}\, \kappa \, \partial_w. \]
Note that for $a=0$ we get the formulae for the Schwarzschild case.

The conformal geodesic equations (\ref{congeo-component}) can be solved numerically with the above initial data. The behaviour of the curves is similar to the Schwarzschild case, with the notable difference that Cauchy horizons exist. Conformal geodesics pass through the horizons where one needs to change the coordinate representation of the metric to integrate the equations further. It would be interesting to see effects of this behaviour in a Cauchy evolution. 
\section{Reduced conformal field equations}
The equations are written for the following variables: a frame field $e^{\ \mu}_{k}$, a Weyl connection $\hat{\Gamma}^{\ k}_{i \ j}$, the Schouten tensor $\hat{L}_{ij} = \frac{1}{2}\hat{R}_{(ij)} -\frac{1}{4}\hat{R}_{[ij]} - \frac{1}{12}\hat{R}\ g_{ij}$, and the rescaled conformal Weyl tensor $W^{i}_{\ jkl} = \frac{1}{\theta}\,C^{i}_{\ jkl}$. They are 
\begin{eqnarray*}
  \partial_{t} e^{\ \mu}_{k} &=& - \hat{\Gamma}^{\ l}_{k \ 0}e^{\ \mu}_{l}, \\
  \partial_{t} \hat{\Gamma}^{\ k}_{i \ j} &=& - \hat{\Gamma}^{\ k}_{l \ j}\hat{\Gamma}^{\ l}_{i \ 0} + g^{k}_{\ 0} \hat{L}_{ij} + g^{k}_{\ j} \hat{L}_{0i} - g_{j0} \hat{L}^{\ k}_{i} + \theta\, W^{k}_{\ j0i}, \\
  \partial_{t} \hat{L}_{ij} &=& - \hat{\Gamma}^{\ k}_{i \ 0} \hat{L}_{kj} + d_{l} W^{l}_{\ j0i}, \\
  \nabla_{l} W^{l}_{\ ijk} &=& 0.
\end{eqnarray*}
The functions $\theta$ and $d_k$ are known a priori in terms of initial data. Main advantages of this system, especially for numerical work, are the following:  
\begin{itemize}
\item The system consists mainly of ordinary differential equations except the Bianchi equation, which has symmetric hyperbolic reductions. This is advantageous especially when one is dealing with complicated geometries.
\item The location of the conformal boundary is known explicitly.
\end{itemize}
One difficulty is that the equations become degenerate at the set $\mathcal{I}^+$ where null infinity meets spatial infinity \cite{Friedrich98}. A first step in numerical work with these equations is the spherically symmetric case. We know that the conformal Gauss gauge covers the complete Schwarzschild-Kruskal solution and is robust enough to do numerical calculations. Therefore, the Cauchy problem in spherical symmetry with Schwarzschild initial data can be expected to be solvable without difficulties.
\subsection{Spherical symmetry}
 To determine the only non-vanishing Weyl tensor component in spherical symmetry, we need to solve the following subsystem of the reduced conformal field equations
\begin{eqnarray} \label{cauchy}
\partial_\tau \Gamma^{\ 2}_{2 \ 0} &=& - (\Gamma^{\ 2}_{2 \ 0})^2 + \hat{L}_{22} - \frac{1}{2}\theta \, W^0_{\ 101}, \nonumber\\
\partial_\tau \hat{L}_{22} &=&  - \hat{\Gamma}^{\ 2}_{2 \ 0}\, \hat{L}_{22} - \frac{1}{2} d_0\, W^0_{\ 101}, \nonumber \\
\partial_\tau W^{0}_{\ 101} &=& - 3 \, \Gamma^{\ 2}_{2 \ 0} \, W^0_{\ 101}. 
\end{eqnarray}
The quantities  $\theta$ and $d_0$ are given by
\begin{eqnarray*} \theta(\tau) = \frac{\Omega}{\kappa}\left(1 - \tau^2 \frac{\kappa^2}{\omega^2}\right), \qquad  d_0(\tau) = -2 \tau \frac{\kappa \Omega}{\omega^2}. \end{eqnarray*}
The degeneracy of the equations at $\mathcal{I}^+$ is not present in the simple case of spherical symmetry. The reduced conformal field equations become a system of ordinary differential equations. 

The initial data can be calculated as in \cite{Friedrich98}. We choose a frame such that $h(\bar{e}_a, \bar{e}_b) = \delta_{ab}$ with respect to $h=\Omega^2 \tilde{h}$ and then rescale it via $e^{\ \mu}_{a} = \kappa \bar{e}^{\ \mu}_{a}$. The connection coefficients are calculated with respect to $e_a^{\ \mu}$. For the Schouten tensor and the electric part of the Weyl tensor we calculate
\[ L_{ij} = \kappa^2 \left( \frac{1}{\Omega} \,t_{ij} + \frac{1}{12}\, \bar{r} \, \bar{h}_{ij} \right), \qquad W^0_{\ i0j} = \kappa^3 \left( \frac{1}{\Omega^2} \, t_{ij} + \frac{1}{\Omega}\, s_{ij} \right) \]
with $t_{ij} = \bar{D}_i \bar{D}_j \Omega - \frac{1}{3}\,\delta_{ij} \, \bar{D}_k \bar{D}_l \Omega \ \delta^{kl}$ and  $s_{ij} = \bar{r}_{ij} - \frac{1}{3} \bar{r} \delta_{ij} $ are calculated with respect to $\bar{e}_a^{\ \mu}$. 
By solving the system (\ref{cauchy}) we can generate the entire Schwarzschild-Kruskal solution including the conformal extension.

To check the quality of the solution, we compare the Weyl tensor in the frame representation adapted to the conformal geodesics. In the conformal Gauss gauge, we calculate the timelike coframe using the orthonormality conditions
\[  \sigma^0_{\ 0} = \frac{e_1^{\ 1}}{e_1^{\ 1}e_0^{\ 0} - e_0^{\ 1}e_1^{\ 0}}, \ \qquad \ \sigma^0_{\ 1} = - \frac{e_1^{\ 0}}{e_1^{\ 1}e_0^{\ 0} - e_0^{\ 1}e_1^{\ 0}}  \]
and compare the rescaled Weyl tensor in the conformal Gauss gauge on the Schwarzschild-Kruskal background given by
\begin{equation}
W^0_{\ 101} = \sigma^0_{\ \mu} e^{\ \nu}_{1} e^{\ \lambda}_{0} e^{\ \rho}_{1}\  \frac{1}{\theta}C^\mu_{\ \nu\lambda\rho}, 
\end{equation}
with the result of the numerical simulation of the Schwarzschild initial data. The calculations agree with each other within the numerical order of integration, so the resulting spacetime can be illustrated by the same conformal diagram Fig.~\ref{fig:conf_geodesics}

While numerical tests of tetrad formulations in spherical symmetry can be quite delicate and instructive \cite{Buchman05}, the simplicity of (\ref{cauchy}) does not allow us to draw representative conclusions for the general case. Still, the study suggests that global numerical calculations for spacetimes with less symmetry might be possible using the reduced conformal field equations. 
\section{Conclusions}
Basic motivations for numerical studies of conformal compactification are to avoid artificial outer boundaries and to have rigorous analysis tools for numerically generated spacetimes with efficient codes. A detailed understanding and numerical control of the conformal structure at spatial infinity is required for a successful implementation of conformal compactification. The tool for this task, i.e. the reduced conformal field equations, has been developed by Friedrich. Within this article we have studied the numerical feasibility of the underlying conformal Gauss gauge in simple cases. 

The results of this study can be summarized as follows: One can numerically reproduce Friedrich's covering of the complete Schwarzschild-Kruskal solution using the conformal Gauss gauge. Numerical simulations suggest that one can also cover the complete Kerr solution. For the first time, the Cauchy problem for the simplest asymptotically flat black hole spacetime has been solved globally including spacelike, null and timelike infinity and the domain close to the singularity. This was made possible by the reduced conformal field equations.

As the studied examples are very special, they are not representative for the general case. One should study the robustness of the conformal Gauss gauge in radiative spacetimes. My current work is directed towards numerical simulation of solutions to the reduced conformal field equations with a non-vanishing radiation field along null infinity. The imposed geometry by the cylinder at spatial infinity and the degeneracy of the equations at $\mathcal{I}^+$ requires highly developed numerical techniques for the solution of this problem.
\ack
I thank Helmut Friedrich for suggesting the subject, clarifying many questions and for a careful reading of the manuscript. I thank Sascha Husa for helping me with computational problems. Thanks also to Florian Beyer for making suggestions on the manuscript.
This work was supported by the SFB/Transregio 7 ``Gravitational Wave Astronomy'' of the German Science Foundation.

\end{document}